\documentclass[preprint]{aastex}
\slugcomment{submitted to {\it The Astronomical Journal}}
\begin{document}
\title{$vbyCa$H$\beta$ CCD Photometry of Clusters. VI. The Metal-Deficient Open Cluster NGC 2420}
\author{Barbara J. Anthony-Twarog, Delora Tanner, Misty Cracraft, and Bruce A. Twarog}
\affil{Department of Physics and Astronomy, University of Kansas, Lawrence, KS 66045-7582}
\affil{Electronic mail: bjat@ku.edu,cracraft@stsci.edu, btwarog@ku.edu}

\begin{abstract}
CCD photometry on the intermediate-band $vbyCa$H$\beta$ system is presented for the metal-deficient open cluster, NGC 2420. Restricting the data to probable single members of the cluster using the CMD and the photometric indices alone generates a sample of 106 stars at the cluster turnoff. The average $E(b-y)$ = 0.035 $\pm$0.003 (s.e.m.) or $E(B-V)$ = 0.050 $\pm$0.004 (s.e.m.), where the errors refer to internal errors alone. With this reddening, [Fe/H] is derived from  both $m_1$ and $hk$, using $b-y$ and H$\beta$ as the temperature index. The agreement among the four approaches is reasonable, leading to a final weighted average of [Fe/H] = --0.37 $\pm$0.05 (s.e.m.) for the cluster, on a scale where the Hyades has [Fe/H] = +0.12. When combined with the 
abundances from DDO photometry and from recalibrated low-resolution spectroscopy, the mean metallicity becomes [Fe/H] = --0.32 $\pm$0.03. It is also demonstrated that the average cluster abundances based upon either DDO data or low-resolution  spectroscopy are consistently reliable to 0.05 dex or better, contrary to published attempts to establish an open cluster metallicity scale using simplistic offset corrections among different surveys. 
\end{abstract}

\section{INTRODUCTION}
This is the sixth paper in an extended series detailing the derivation of fundamental parameters in star clusters using precise intermediate-band photometry to identify probable cluster members and to calculate the cluster's reddening, metallicity, distance and age. The initial motivation for this study was provided by \citet{tat97}(hereafter TAAT), who used a homogeneous open cluster sample to identify structure within the galactic abundance gradient. The open clusters appear to populate a bimodal abundance distribution with the clusters interior to a galactocentric distance of $\sim$10 kpc averaging [Fe/H] $\sim$ 0.0 while those beyond this boundary average [Fe/H] $\sim -0.3$. The dispersion about each peak is less than 0.1 dex and implies that within each zone, no statistically significant abundance gradient exists, in contrast with the traditional assumption of a linear gradient over the entire disk. This structure has since been corroborated through the use of Cepheids by \citet{an12,an02,lu03} and, most recently, with OB stars \citep{da}. Potential evidence for the lack of  a simple linear gradient across the entire galactic disk, particularly in the outer portion, is found in the comparative discussion of NGC 2243 and Berkeley 29 in \citet{at05}, as well as the in the color-magnitude diagram (CMD) analysis of Berkeley 22 by \citet{di05} and the spectroscopic cluster work of \citet{yo05}. The exact origin and reason for the survival of the discontinuity remain unresolved issues, though delayed evolution of the outer disk through mergers \citep{tat97, yo05} and the radial variation in the impact of spiral structure on star formation and stellar dynamics appear to be potential contributing factors \citep{sc01,mi02,lp03}. 

For balance, it should be noted that not all analyses agree with the interpretation that the abundance gradient of the disk exhibits discontinuous structure rather than a simple linear relation. In some cases, a linear relation is selected because the investigation has high quality data, but of insufficient quantity to statistically validate the discontinuity \citep{fj02}, while others have a large sample, but with abundances of insufficient accuracy to look for fine structure \citep{ch03}. Investigations that merely dismiss the reality of the feature (see, e.g. \citet{ca98,gr00}) invariably use inhomogeneous and outdated material that obscures structure behind unnecessary scatter. We will return to this issue in Section 5. 

Detailed justifications of the program and the observational approach adopted have been given in previous papers in the series \citep{AT00a,AT00b,TW03,at04,
at05} (hereafter referred to as Papers I through V) and will not be repeated. Suffice it to say that the reality of the galactic features under discussion will remain questionable unless the error bars on the data are reduced to a level smaller than the size of the effect being evaluated or the size of the sample is statistically enhanced. The overall goal of this project is to do both.

An equally important aspect of this research is detailed testing of stellar evolution theory as exemplified by comparisons to stellar isochrones based upon models derived under a variety of assumptions. The agreement with (or the deviation from) the predicted distribution and location of stars within the CMD has consistently provided a lever for adjusting our degree of confidence in the specifics of stellar interiors as a function of mass and age. A valuable illustration of this can be found in the discussion of the red giant branch distribution (first-ascent and clump giants) in NGC 3680, in comparison to NGC 752 and IC 4651, clusters of comparable age (Paper IV).

The focus of this paper is the metal-deficient open cluster, NGC 2420. Since the high-quality photometric work of \citet{mfg}, this cluster has served as the definitive example of the older, moderately metal-deficient open clusters beyond the solar circle. Its intermediate status between the solar-metallicity open clusters near the sun and the clearly metal-deficient population of globulars tagged it early on as a potential transition object between the two populations, if not a system that overlapped with the metal-rich tail of the globular clusters. Discrepancies between the predicted and the apparent relative positions of the key CMD features during comparisons with metal-rich globulars such as 47 Tuc \citep{dm} generated ongoing debate about the true metallicity of both clusters \citep{pil,coh}. With the addition of new photometric and spectroscopic data in the 80's and 90's \citep{ss87}, as well as the exquisitely defined CMD based upon new CCD data \citep{aks}, an [Fe/H] between --0.3 and --0.5, depending on which scale one adopted, appeared secure. 
Recently, however, preliminary work by \citet{del} based upon high precision CCD broad-band $(UBVRI)$ photometry has suggested the possibility of a higher metallicity.
Since this evidence is tied to observations of the cluster turnoff and unevolved main sequence while virtually all current abundance estimates for the cluster are based upon observations of the giants, the remote possibility remains that the evolved stars might suggest
a distorted view of the true [Fe/H] for the cluster because of some unknown evolutionary effect or simply an intrinsic error in the common data used to analyze and interpret the cluster giants. 

With access to the northern hemisphere via the WIYN 0.9m at KPNO, intermediate-band $(vbyCa$H$\beta)$ observations seemed an ideal means of directly measuring the intrinsic properties of the cluster using the same stars studied by \citet{del}, while adding this crucial cluster to the network of objects studied to date for use in defining the abundance structure of the disk. In addition, because NGC 2420 was adopted as a standard cluster in setting the metallicity scale for \citet{fj02}, an independent metallicity estimate allows us to check the offset and dispersion in the transformation between the DDO and spectroscopic metallicity scales adopted in the merged sample of TAAT.

Section 2 contains the details of the $vbyCa$H$\beta$ CCD observations, their reduction and transformation to the standard system, and a search for photometrically anomalous stars. In Sec. 3 we discuss the CMD and begin the process of identifying the sample of probable cluster members. Sec. 4 contains the derivation of the fundamental cluster parameters of reddening and metallicity. Sec. 5 includes a discussion of our [Fe/H] in the context of past estimates and an analysis of the overall reliability of the DDO and low-dispersion spectroscopic abundances. Sec. 6 summarizes our conclusions regarding NGC 2420 and the status of temporal and spatial structure within the abundance gradient of the galactic disk. The results of the current discussion supercede preliminary results reported in \citet{act04}.

\section{THE DATA}

\subsection{Observations: Photoelectric $Caby$}
The $Caby$ system was initially defined and developed using traditional photoelectric photometry obtained at a number of observatories, but predominantly CTIO and KPNO, between 1983 and 1996. These observations, primarily of field stars, have been discussed and analyzed in a series of papers \citep{att91, tat95, at00} culminating most recently in the definition of the system for Hyades main sequence stars \citep{at02}. In addition to the Hyades, a number of stars were observed in open and globular clusters with the intent of providing internal standards for future CCD work. Included in this sample were 2 stars in the field of NGC 2420, observed with a pulse-counting system equipped with an S-20 photomultiplier on the 1.5 m telescope at CTIO in 1990. The cluster stars were transformed and reduced with the field stars and should be well tied to the standard system. Details on the reduction and merger of the photometry over a series of nights and runs may be found in \citet{tat95} and will not be repeated here. Suffice it to say that the $V$ and $(b-y)$ values are on the system of \citet{ols93}. The identification numbers are on the system of the WEBDA data base. The results of the single observations for the two stars are as follows: for star 115, the $V$, $b-y$, and $hk$ numbers are 11.546, 0.690, and 1.137, respectively, while the analogous data for star 140 are 11.487, 0.763, and 1.417. Given that these are single observations, estimates of the likely errors must be based upon stars of comparable magnitude and color observed multiple times over the same runs. A rough approximation is that $b-y$ should be good to $\pm$0.010, while $V$ and $hk$ should have errors closer to $\pm$0.015. The stars will not be used in the calibration of the CCD photometry but will serve, instead, as a check on the reliability of the procedure for calibrating the red giants on the $Caby$ CCD system.

\subsection{Observations: CCD $vbyCa$H$\beta$}
The new photometric data for NGC 2420 were obtained using the 
0.9-meter WIYN telescope on Kitt Peak.  The S2KB detector is a $2048 \times 2048$ CCD mounted at the the $f/7.5$ focus of the telescope.  The field size is approximately $20'$ on a side.  We obtained frames of NGC 2420 over two six-night runs in January 2003 and January 2004, using our own $3'' \times 3''$ H$\beta$ and $Ca$ filters and borrowed $4'' \times 4''$ $y,b,$ and $v$ filters from KPNO.  No suitable Str\"omgren $u$ filter was available.  Processing of all frames through bias subtraction, trimming and flat fielding was accomplished at the University of Kansas using standard IRAF routines. Dome flats were used for all filters.  

Data from five nights in 2003 and 4 nights in 2004 were incorporated in the analysis of NGC 2420 photometry.  For each of our six filters, 11 to 15 frames were analyzed as detailed below.  The total exposure times represented in our analyzed photometry are 35, 34, 34, 104, 46 and 81 minutes for the $y, b, v, Ca$, H$\beta$ wide and narrow filters, respectively.

\subsection{Reduction and Transformation}
Previous papers in this series describe the procedures used to produce high precision, accurately calibrated photometry from CCD data. IRAF Allstar routines are used to obtain complete sets of profile-fit magnitudes for all stars on every program frame.  Paper I, in particular, provides a comprehensive description of the steps used to produce a set of average indices of high internal precision for the stars in the program cluster.
 
Regardless of the internal precision achieved by averaging large numbers of frames, the accuracy of the photometric calibration is limited by other factors, including the breadth of parameter space covered by standard stars and observational conditions. Our approach for these steps is described extensively in Paper IV with a brief delineation presented here for NGC 2420's calibration.

Standard stars in the field and in clusters, as well as uncrowded stars in the program cluster, are observed on photometric nights and reduced using a consistent aperture measurement and airmass correction strategy.  Seeing for the 2003-2004 runs at KPNO-WIYN was such that fairly large aperture radii of 11 to 14 pixels were used with sky annuli set to yield comparable areas.  This approach permits the standardization of aperture photometry in the program cluster obtained on any photometric night. By extension, we are able to apply calibration relations to the high precision profile-fit photometric indices by determining the mean difference between the calibrated aperture photometry for a photometric night's data in the program cluster and the profile-fit indices. In a rich field such as NGC 2420, it is possible to determine the mean difference between the averaged profile-fit indices and indices from a photometric night to precisions of 0.01 mag or less.  This essentially establishes an {\it aperture-correction} to apply to the psf-based photometry.

Our sources for standard stars include datasets in clusters and field star catalogs.  For uniform $V$ magnitudes, $(b-y)$ colors and $hk$ indices, our primary source is our own 1995 catalog \citep{tat95}.   For F and G stars, observational catalogs compiled by \citet{ols83,ols93,ols94} are used; these catalogs are particularly important for developing separate calibration relations for different luminosity and color classes of stars. Additional photoelectric $hk$ observations in NGC 752 have been obtained at Mt. Laguna Observatory in 1999 and will be described more fully in a separate publication.  NGC 752 is already a rich source of standard stars with Str\"omgren and H$\beta$ indices from \citet{cb,tw752,ni88}.  Str\"omgren and H$\beta$ data in M67  \citep{ntc} complete our standard star reservoir.  

Discussion of the calibration of the H$\beta$ indices may illustrate the steps taken to calibrate photometric indices in NGC 2420.  Three nights of the 2003-2004 CCD imaging runs were photometric. For each of these nights, from four to twelve field star standards and between 15 and 30 additional standard stars in the cluster fields were observed with H$\beta$ filters, as well as H$\beta$ frames obtained in the NGC 2420 field.  All of the indices for the standard stars were compared to standard values to determine a common slope for the calibration equation.  A separate zeropoint for the calibration equation was determined for each photometric night. The standard deviations for these linear fits are typically 0.01 magnitude, with the standard error of the mean for the zeropoint value consequently much smaller.  A direct comparison was made between the aperture-based H$\beta$ indices in NGC 2420 from each photometric night  to the averaged psf-based H$\beta$ indices to determine the mean offset between the data sets. Typically, 10 to 20 uncrowded stars are common to both sets of indices, enabling an aperture correction to be determined with a standard error of the mean below 0.006 mag. Based on the calibration equations derived from the photometric nights, a single calibration equation is developed to transform the psf-based H$\beta$ indices to the standard system. The zeropoint for that equation is a weighted average of the zeropoints developed from each photometric night, with the aperture correction and the standard star data contributing in quadrature to the standard error of the mean for the zeropoint. The calibration equation describing the H$\beta$ photometry in NGC 2420 has a standard error of the mean for the zeropoint of 0.004 mag.

Similar approaches are used to develop calibration equations to relate instrumental indices to standard $V$ magnitudes, $b-y$, $m_1$ and $hk$ color indices, with separate calibration equations derived for cooler evolved stars' $m_1$ indices. For most indices, the calibration equation benefits from having at least two photometric nights contribute to the accuracy of the zeropoint. The exception is the calibration equation for red giant $m_1$ values, for which only one photometric night had sufficient standards to constrain the zeropoint. The standard errors of the mean for the zeropoints of the calibration equations for V, $(b-y)$, $m_1$ (red giants), $m_1$ (dwarfs), and $hk$ are 0.003, 0.003, 0.004, 0.006, and 0.009, respectively. The final photometry for all stars with at least two observations in $y$ and $b$ is given in Table 1. Identifications and XY coordinates are on the system found in the WEBDA Cluster Data Base. For those stars not included in the WEBDA database, an identification number has been created beginning with 6000. Figures 1 and 2 illustrate the typical errors in the photometric indices as a function of $V$ magnitude.

\subsection{Potential Variables}
The photometry discussed above was obtained to reliably define the cluster in intermediate-band CMD and color-color diagrams so, while over 100 frames have been analyzed, the distribution is approximately 12 frames per filter at the bright end and less than half that among the fainter stars. Despite the lack of optimization for detecting variable stars, it should be possible to identify at least some candidates with modest to large amplitude variations over a range of timescales. Comparison with the surveys designed specifically to identify such anomalies \citep{kim,ka} should provide an independent test of our results.

The obvious criterion for discovering variables is a large photometric scatter in the indices for a star relative to what is expected at a given magnitude. The points that deviate from the well-defined trends in Figs. 1 and 2 should generate a first cut for our sample. However, deviant points could exist in Figs. 1 and 2 for a variety of reasons, most of which have nothing to do with variability. Since we are plotting the standard error of the mean, two stars at the same magnitude level with identical frame-to-frame photometric errors will populate different locations in the figure because one star only appears on a small subset of the frames. The lack of a complete set of magnitudes is a common occurrence for stars in the outer edges of the cluster field, as the cluster position on the CCD chip is shifted from night to night and run to run to avoid placing the same stars on the same bad columns and pixels, and to extend the field coverage beyond the minimum allowed by one CCD frame. Additionally, a paucity of measures for a star in any region of the cluster may be a sign of image overlap/confusion, resulting in poor or inadequate photometry. Finally, while the frames have been processed using a variable point-spread-function, the success of such modifications may decline as one nears the edge of the chip and the photometric scatter in any filter may be expected to increase in the outer zone of the field, irrespective of the number of frames included in the average.

To tag possible variables, the stars that deviated the most from the mean relations for $V$ and $b-y$ in Fig. 1 for $V$ brighter than 17.40 were identified and the errors renormalized to a uniform number of frames. If the renormalization placed the star near the standard relation, it was excluded. Stars located more than 700 pixels from the cluster center were excluded. Stars with fewer than 2/3 of the expected frames at a given magnitude were checked for possible contamination/confusion with a neighbor and eliminated if a companion was found. Finally, the individual magnitude errors were checked and a star retained if it exhibited large scatter in $y$ and at least one additional filter among $b$, $v$, and $Ca$, and showed a larger than expected error in H$\beta$. The final result was the identification of only two possible variables within the sample: 186 and 6209. These stars are noted by filled symbols in Figs. 1 and 2. Of the two stars, 186 is a probable non-member \citep{aks}. 

What is surprising about the paucity of identified variables is the agreement with the work of \citet{ka} and \citet{kim}, who find no evidence for variability among any of the stars they have surveyed in the cluster. While the timescale over which variability was monitored amounts to hours, the anomalously low detection rate for a cluster rich in binaries deserves further attention, as noted by \citet{kim}. 
The significance of this result of this result remains an unknown due to the lack of comprehensive variability data on other clusters of comparable age and metallicity. However, \citet{kim} also studied NGC 2506 at the same time as NGC 2420. NGC 2506 is of comparable age and metallicity and appears to have an equally high percentage of binaries based upon the distribution in the CMD \citep{mtf}. Using a comparable number of CCD frames on a similar number of stars over a longer timeline, \citet{kim} discovered three $\delta$ Sct stars and one eclipsing binary in NGC 2506.
An alternative means of detecting potential longer period systems is through the comparison of our $V$ photometry with comparable high quality surveys of the past.

 \subsection{Comparison to Previous Photometry: Intermediate-Band Data}
The first check on the CCD photometry is a comparison with the photoelectric data discussed at the start of this section. For the two red giants, the average residuals, in the sense (PE - CCD), for $V$, $b-y$, and $hk$ are --0.014 $\pm$0.009, --0.013 $\pm$0.002, and +0.012 $\pm$0.005, respectively, where the errors quoted are simply one-half the difference between the two residuals. The reasonable agreement between the systems is within the expected errors, an encouraging result, but only applicable to the calibration for the cooler, evolved stars.

Additional photoelectric observations on the $uvby$ system are available from two sources, \citet{eg78} and \citet{ms81}. The former reference covers a larger sample of stars, including turnoff stars and bright giants, while the latter includes only 2 bright blue stragglers. The data of \citet{ms81} are most likely to resemble the standard system, so we will deal with them first. For the two stars, the mean residuals in $V$, $b-y$, and $m_1$, in the sense (PE - CCD), are --0.095 $\pm$0.004, +0.022 $\pm$0.006, and +0.003 $\pm$0.007, respectively, with the errors defined as above. There is a clear systematic error in the $V$ system of \citet{ms81}, but the differences in $b-y$ and particularly $m_1$ are modest and internally consistent.

Turning to the data of \citet{eg78}, the average residuals for all 10 stars are +0.039 $\pm$0.029, +0.014 $\pm$0.028, and --0.010 $\pm$0.040 for $V$, $b-y$, and $m_1$, respectively, where the error quoted is the standard deviation for a single residual. The sample includes one observation of one star at $V$ $\sim$15.5, a magnitude fainter than the next brightest star and a difficult observational photoelectric challenge with even a 1.5m telescope. If we drop this star, the sequence of  average residuals becomes +0.033 $\pm$0.024, +0.008 $\pm$0.022, and --0.013 $\pm$0.041.

However, it is also known \citep{eg76} that for cooler stars, the photometric $uvby$ system devised by Eggen is not that of \citet{ols93}, the
standard adopted here. If we restrict the comparison to the 5 stars at the turnoff of NGC 2420, the mean residuals in $V$, $b-y$, and $m_1$ are +0.034 $\pm$0.036, +0.007 $\pm$0.039, and  --0.001$\pm$0.033, respectively. Finally, the faintest star noted above is within this group. If we exclude it from the comparison, the last set of comparisons from 4 stars becomes +0.020 $\pm$0.018, --0.009 $\pm$0.020, --0.005 $\pm$0.036.

\subsection{Comparison to Previous Photometry: Broad-Band $V$}
While there have been a number of photographic studies of the cluster, particularly \citet{mfg} and \citet{mnb}, we will focus on two sources of high-quality CCD data, \citet{aks} and \citet{st00}. As in the previous papers in the series, we will compare the residuals in the photometry for the entire sample and for the brighter stars alone since the reliability of the latter sample is more relevant to the determination of the cluster parameters. Our interest in the residuals is twofold: to measure the photometric uncertainties in comparison with what is expected from the internal errors and to identify stars that exhibit significant deviations from one study to another. The deviants will be a mixture of misidentifications, bad photometry, and, of particular value, long-term variables or longer-period eclipsing binaries that will not be readily exposed by the variable star searches done to date. Data for the published surveys have been taken from the WEBDA Cluster Data Base. The results are summarized in Table 2.

For 531 stars at all $V$, the residuals, in the sense (Table 1 - REF), for the comparison with \citet{aks} average $0.001 \pm 0.034$. If 10 stars with deviations greater than 0.10 mag are eliminated, the remaining 521 stars average $+0.001 \pm 0.030$; the deviant stars are all fainter than $V = 18.0$. No color term of significance, based upon the low correlation coefficients for the attempted fits (R below 0.10), is found among the residuals. If we restrict the sample to 229 stars brighter than $V = 16.5$, the average residual becomes $+0.005 \pm 0.022$. Dropping the only star with a residual greater than 0.075 mag (star 54), the average becomes $+0.005 \pm 0.021$, with weak evidence for a color term (R = 0.38).

Turning to the data for \citet{st00}, there are 177 stars at all magnitudes, none of which exhibits residuals greater than 0.10 mag. The average residual is $+0.007 \pm 0.027$, with stronger evidence for a color term with a slope of $-0.111$ (R = 0.56).  Turning to the 114 stars brighter than $V$ = 16.5, the average residual is $+0.017 \pm 0.023$; removing the one deviant (3335) with a residual larger than 0.075 mag, leaves an average $V$ of $+0.017 \pm 0.021$ for 113 stars.

With the exception of the brighter stars of \citet{st00}, the $V$ magnitude scales for all three studies appear to be within 0.01 mag of each other. Again, what is surprising to note is the almost total absence of variability among data sets obtained over a time frame close to a decade. Of the two deviants brighter than $V$ = 16.5, star 54 is part of an optical double that provides a logical explanation for the discrepancy, while 3335 is a highly probable field star sitting between the main sequence and the giant branch, well above the range of a binary system.

\section{The Color-Magnitude Diagram: Thinning the Herd}

Membership information of variable quality is available for many of the brighter stars in the cluster down to just below the turnoff \citep{aks}. Fortunately, the exceptionally well-defined nature of the $BV$ CMD \citep{aks} will allow us, in conjunction with restrictions to the cluster core region, to eliminate with high probability most of the cluster nonmembers. Thus NGC 2420 will be treated as a program cluster analogous to NGC 2243 (Paper V).

The CMD for all stars with two or more observations in both $b$ and $y$ is shown in Fig. 3. Stars with errors greater than $\pm$0.010 in $b-y$ are presented as crosses. Though not as tightly defined as the more restricted sample of \citet{aks}, the primary features of the CMD extending four magnitudes below the turnoff are readily apparent, as is the rich clump of stars at $V$$\sim$12.5. The morphology of the CMD clearly is indicative of a cluster intermediate in age between NGC 752 \citep{da94} and M67 \citep{san}. The small internal errors in the photometry are maintained to $V$$\sim$17 or, at least, $b-y$ $\leq$ 0.4. 

To reduce field star contamination, we next restrict the data to the spatial core of the 
cluster. The radial distribution of NGC 2420 has been discussed in detail by \citet{leo} and \citet{aks}. We applied the same technique used in past papers in this series to define the probable center of the cluster and identified a position similar to that found by \citet{leo}, not unexpected given the wider spatial extent of that study compared to \citet{aks}. Stars within 300 WEBDA coordinate units (4.5\arcmin) of the newly defined coordinate system were retained; this provided an area only slightly larger than but mostly inclusive of the stars studied in \citet{aks}, allowing us to use any information presented for these stars in that study. The improvement in the homogeneity of the sample is illustrated by Fig. 4, where the core stars are plotted using the same symbols as in Fig. 3. One could eliminate more of the probable field stars by restricting the spatial range even further, but the zone of the CMD that will be used to derive the key cluster parameters can be refined in alternative ways without the need for wholesale elimination of a large number of true members.

\subsection{Thinning the Herd: CMD Deviants}
For purposes of optimizing the derivation of the cluster reddening and metallicity, our interest lies in using only single stars that evolve along a traditional evolutionary track and have indices in a color range where the intrinsic photometric relations are well defined. 
With this in mind, we further reduce the sample by including only stars with errors in $b-y$ of 0.010 or less, in the magnitude 
range from $V = 14.2$ to 17.0 and the color range from $b-y \leq 0.40$, leaving 163 stars.

Given the expected dominance of the cluster sample over the field stars, it is probable that the majority of stars at the cluster turnoff are members, though not necessarily single stars. As in past papers in the series, it is valuable to identify and remove the likely binary systems from the sample to avoid distortions of the indices through either anomalous evolution or simple photometric combinations of the light for stars in significantly different states. For stars in the vertical portion of the turnoff, separation of the stars into two parallel sequences, one single and one composed of binaries, becomes a challenge because the single stars evolving away from the main sequence produce an evolutionary track that curves toward and crosses the rich binary sequence composed of unevolved pairs. Moreover, despite the high internal precision of the $b-y$ indices, the scatter near the turnoff is unexpectedly large, in contrast with the broad-band data of \citet{aks}.

In our last two papers, we have enhanced our ability to delineate the single and binary sequences through the use of $v-y$, a color index with a larger baseline in wavelength and greater sensitivity to temperature change than $b-y$, as championed by \citet{me00}. The greater baseline gives the index greater temperature sensitivity but, since $v$ is dominated by metallicity effects rather than surface gravity and all the stars within the cluster supposedly have the same [Fe/H], the merger of the two sequences due to evolutionary effects among the single stars should be less of an issue than the use of $u-y$.  For the 163 stars remaining after our previous cuts, the $V, v-y$ diagram is illustrated in Fig. 5. For stars redder than $v-y$ = 0.72, the main sequence is displayed as a band 0.75 mag wide in the vertical direction, exactly as expected for a range of pairs with the upper bound generated by two identical stars. Using this figure, we did an initial identification of the deviants by tagging as potential binaries and/or nonmember interlopers any star more than 0.02 mag redder than the narrowly defined and, implicitly, unevolved main sequence. One star was identified as a probable field star due to its location blueward of the main sequence.

To test whether this classification makes any sense, we have plotted the same stars with the same symbols in a $V, (Ca-y)$ diagram, as illustrated in Fig. 6. The location of the $Ca$ filter in the near ultraviolet acts as a proxy for the even bluer $u$ filter, without the issue of surface gravity dependence. The correspondance between the two figures is excellent. Of the 28 stars initially identified as deviant in Fig. 5, 25 were classified as such in Fig. 6. Two additional deviants were classified using the positions of the stars in Fig. 6, bringing the total to 30 (filled red circles). It should be emphasized that the significant improvement in the delineation of the unevolved main sequence as one moves from $b-y$ to $v-y$ and $Ca-y$ is mostly a result of the increasing baseline between the filters. However, as mentioned earlier, the scatter in $b-y$ is larger than expected and appears to have its origin within the $b$ magnitudes. Whether this effect is a product of unexplained photometric errors or an indication of something intrinsic to the stars remains unknown.  
The range in color among the select stars in the CMD is 0.13 mag, 0.30 mag, and 0.50 mag for $b-y$, $v-y$, and $Ca-y$, respectively.

As a final check, in Fig. 7 we plot the broad-band CMD for the same stars using the CCD data of \citet{aks}. The sample in Fig. 7 includes 155 stars because 8 stars fell outside the area of the broad-band survey. As is apparent, the ability to identify stars that deviate from the normal, single-star main sequence using either $v-y$ or $Ca-y$ is beautifully confirmed. Only one star remains superposed on the main sequence in the $BV$ data, one of the few stars with contradictory positions in $v-y$ and $Ca-y$. For consistency with past papers, all 30 stars classed as deviants in either diagram will be excluded, leaving 133 potential stars for photometric analysis. 

\section{FUNDAMENTAL PROPERTIES: REDDENING AND METALLICITY}

\subsection{Reddening}

All 133 stars have multiple measures in every filter. However, to minimize the impact of photometric errors, we exclude 26 stars with errors in H$\beta$ above 0.010, and one star with an error in $m_1$ above 0.014. The remaining 106 stars have errors in both $m_1$ and $hk$ less than or equal to 0.014. The average standard errors of the mean for the various indices are 0.0057, 0.0084, 0.0084, and 0.0066 for $b-y, m_1, hk$, and H$\beta$, respectively. 

As discussed in Paper I, derivation of the reddening from intermediate-band photometry is a straightforward, iterative process given reliable estimates of H$\beta$ for each star. Since metallicity and reddening both affect the intrinsic colors and the evaluation of each parameter, the reddening is derived for a range of assumed values for $\delta$$m_1$($\beta$), the metallicity index, then the metallicity index is derived for a range of assumed reddening values. Only one combination of $E(b-y)$ and $\delta$$m_1$ will simultaneously satisfy both relations. The primary decision is the choice of the standard relation for H$\beta$ versus $b-y$ and the adjustments required to correct for metallicity and evolutionary state. The two most commonly used relations are those of \citet{ols88} and \citet{ni88}. As found in previous papers for IC 4651, NGC 6253, NGC 3680, and NGC 2243,  both produce very similar if not identical results. 

The modest twist in the current investigation comes from the lack of $c_1$ due to the absence of the $u$ filter. For the intrinsic colors, most relations include a term weakly dependent upon $\delta$$c_1$($\beta$), the difference between the observed index at a given value of H$\beta$ and the standard relation for unevolved stars; the more evolved a star is, the larger the value of $\delta$$c_1$($\beta$). Fortunately, because we have removed the most probable sources of confusion between evolution and binarity in the CMD, any deviation from the unevolved main sequence in the CMD for stars in Fig. 7 must be attributed to evolution. As a crude approximation to provide some insight into the effect of $c_1$, we have calculated the position of each star in the CMD above the zero-age main sequence at the same value of H$\beta$, using the coolest stars in the sample to zero the position of the relation defining the unevolved main sequence. The value of $\delta$$M_V$($\beta$) has then been converted to $\delta$$c_1$($\beta$) and thus to $c_1$.  We emphasize that the inclusion of these terms has an exceedingly minor impact on both the reddening and metallicity estimates, so the added scatter created by the possible errors in the $c_1$ estimates is negligible.

Processing the indices for the 106 stars through both relations generates $E(b-y)$ = 0.037 $\pm$0.024 (s.d.) with \citet{ols88} and $E(b-y)$ = 0.034 $\pm$0.022 (s.d.) with \citet{ni88}, with $\delta$$m_1$($\beta$) = 0.046 $\pm$0.002 (s.e.m.). As a compromise, we will take the weighted average of the two and use $E(b-y)$ = 0.035 $\pm$0.003 (s.e.m.) or $E(B-V) = 0.050 \pm 0.004$ (s.e.m.), where the errors refer to internal errors alone, in the analyses that follow.  Inclusion of potential external errors in the definition of the photometric system raises the cumulative error budget to $\pm 0.006$ in $E(b-y)$ and $\pm 0.008$ in $E(B-V)$.

\subsection{Metallicity from $m_1$} 
Given the reddening of $E(b-y)$ = 0.035, the derivation of [Fe/H] from the $m_1$ index is as follows. The $m_1$ index for a star is compared to the standard relation at the same color and the difference between them, adjusted for possible evolutionary effects, is a measure of the relative metallicity. Though the comparison of $m_1$ is often done using $b-y$ as the reference color because it is simpler to observe, the preferred reference index is H$\beta$ due to its insensitivity to both reddening and metallicity. Changing the metallicity of a star will shift its position in the $m_1$,$(b-y)$ diagram diagonally, while moving it solely in the vertical direction in $m_1$, H$\beta$. Moreover, reddening errors do not lead to correlated errors in both $m_1$ and H$\beta$. 

With the exception of Paper V, we have derived the metallicity using $b-y$ and H$\beta$ as the defining temperature index for $m_1$ with, on average, no statistically significant difference in the outcome.  After the publication of Paper IV, alternative [Fe/H] calibrations based upon $b-y$ and $m_1$ were derived by \citet{no04} for F stars and cooler, calibrations that make use of the reddening-corrected indices rather than differentials compared to a standard relation. This approach was first used successfully by \citet{sn89}, but the primary focus of their work was on metal-deficient dwarfs and concerns about the application of the function to solar-metallicity dwarfs limited its adoption for disk stars. 
These concerns proved valid for the metallicity calibration for cooler dwarfs where [Fe/H] was systematically underestimated at the metal-rich end of the scale \citep{tw02}. The more extensive recalibrations for F dwarfs and cooler by \citet{no04} are readily applicable to solar and higher metallicity dwarfs at all colors and eliminate the concerns regarding the original functions of \citet{sn89}. As in Paper V for the another metal-poor open cluster, NGC 2243, we will derive [Fe/H] without reference to H$\beta$ using the \citet{no04} relation and from $\delta$$m_1$(H$\beta$) as in past papers, supplying an independent check of the calibration relations.

After correcting each star for the effect of $E(b-y)$ = 0.035, the mean [Fe/H] using the 
F-star relation of \citet{no04} is --0.488 $\pm$ 0.037 (s.e.m.). In contrast, after correcting each star for the effect of $E(b-y)$ = 0.035 and deriving the differential in $m_1$ relative to the standard relation at the observed H$\beta$,  the average $\delta$$m_1$ for 106 stars is +0.046 $\pm$0.002 (s.e.m.), which translates into [Fe/H] = --0.438 $\pm$0.029 (s.e.m.) for the calibration as defined in \citet{ni88} and adopted in previous papers. As in Paper V, we need to make an additional adjustment. The zero-point of the H$\beta$ metallicity calibration has been fixed to match the adopted value for the Hyades of [Fe/H] = +0.12, i.e., if one processes the data for the Hyades or the standard relation through the [Fe/H] calibration, one is guaranteed to obtain [Fe/H] = +0.12 for any star with $\delta$$m_1$ = 0.000. If the standard relation or the observed data for the Hyades are processed through the \citet{no04} relation, at the cooler end of the scale beyond $b-y$ = 0.32, one obtains [Fe/H] between = +0.12 and +0.16. As $b-y$ decreases, [Fe/H] declines steadily, reaching a minimum near +0.03 near the hotter end of the scale ($b-y$ = 0.23). For the stars in the color range of interest for NGC 2420,  [Fe/H] for the Hyades is systematically underestimated by 0.038 dex relative to the adopted value. Thus, for consistency on the scales,  the [Fe/H] estimate based upon $b-y$ and $m_1$ should be raised to --0.450, the same value, within the errors as the result from the H$\beta$-based relation. Note that the error for an individual estimate from the first approach is larger than the $\delta$$m_1$(H$\beta$) technique because of the enhanced sensitivity of the non-linear terms in the first [Fe/H] calibration to errors in both $b-y$ and $m_1$ and the likelihood of correlated errors between $b-y$ and $m_1$.  Taking into account the external errors in standardizing the photometric indices leads to a cumulative error budget in [Fe/H] of 0.08 dex or less.

The primary weakness of metallicity determination with intermediate-band filters is the sensitivity of [Fe/H] to small changes in $m_1$; the typical slope of the [Fe/H]/$\delta$$m_1$ relation is 12.5. Even with highly reliable photometry, e.g., $m_1$ accurate to $\pm$0.015 for a faint star, the uncertainty in [Fe/H] for an individual star is $\pm$0.19 dex from the scatter in $m_1$ alone. When potential photometric scatter in H$\beta$ and $b-y$ are included, errors at the level of $\pm$0.25 dex are common, becoming even larger for polynomial functions of the type discussed above. In contrast, the effect of errors in the reddening estimate is modest with $\Delta$[Fe/H]/$\Delta$$E(B-V)$ just under 3.
As noted in previous papers in this series, the success of the adopted technique depends upon both high internal accuracy and a large enough sample to bring the standard error of the mean for a cluster down to statistically useful levels, i.e., below $\pm$0.10 dex. Likewise, because of the size of the sample, we can also minimize the impact of individual points such as binaries and/or the remaining nonmembers, though they will clearly add to the dispersion.

\subsection{Metallicity from $hk$}
We now turn to the alternative avenue for metallicity estimation, the $hk$ index. The $hk$ index is based upon the addition of the $Ca$ filter to the traditional Str\"{o}mgren filter set, where the $Ca$ filter is designed to measure the bandpass that includes the H and K lines of Ca II. The design and development of the $Caby$ system have been laid out in a 
series of papers discussing the primary standards \citep{att91}, an extensive 
catalog of field star observations \citep{tat95}, and calibrations for both red giants \citep{att98} and metal-deficient dwarfs \citep{at00}. Though the system was optimally designed to work on metal-poor stars and most of its applications have focused on these stars \citep{atc95,bd96,at00}, early indications that the system retained its metallicity sensitivity for metal-rich F dwarfs have been confirmed by observation of the Hyades and analysis of nearby field stars \citep{at02}. What makes the $hk$ index, defined as $(Ca-b)-(b-y)$, so useful for dwarfs, even at the metal-rich end of the scale, is that it has half the sensitivity of $m_1$ to reddening and approximately twice the sensitivity to metallicity changes. The metallicity calibration for F stars derived in \citet{at02} used $\delta hk$ defined relative to $b-y$ as the temperature index. To minimize the impact of reddening on metallicity, this calibration was redone in Paper III using H$\beta$ as the primary temperature index, leading to the preliminary relation
\medskip
\centerline{[Fe/H]$ = -3.51 \delta hk(\beta) + 0.12$}
\smallskip
with a dispersion of only $\pm$0.09 dex about the mean relation. Though the derived zero-point of the relation was found to be +0.07, it was adjusted to guarantee a Hyades value of +0.12, the same zero-point used for the $m_1$ calibration. Because of the expanded sample of stars with spectroscopic abundances and the revised $m_1$ calibrations for F and G dwarfs by \citet{no04}, a revised and expanded calibration of the $\delta$$hk$ indices, based on both $b-y$ and H$\beta$ is underway \citep{act}.  Modest changes have been generated in the color dependence of the [Fe/H] slope for $\delta$$hk(b-y)$, with even smaller adjustments to the H$\beta$-dependent relation. To ensure that that the metallicities based upon $m_1$ and $hk$ are on the same internal system, we have derived [Fe/H] from the unmodified $m_1, b-y$ function of \citet{no04} for all dwarfs with $hk$ indices and derived linear relations between [Fe/H] and $\delta$$hk(b-y)$ and $\delta$$hk($H$\beta)$ for three different color ranges among the F-stars. 

Applying these modified metallicity calibrations to the $hk$ data for 106 stars
in NGC 2420, with the inclusion of the zero-point adjustment discussed in the previous section, the resulting [Fe/H] values for $hk$ relative to $b-y$ and H$\beta$ are [Fe/H] = --0.341 $\pm$0.020 (s.e.m.) and --0.326 $\pm$0.012 (s.e.m.), respectively. Taking the external errors into account raises the total error budget to just under $\pm$0.04 dex. The dramatic decrease in the scatter with the H$\beta$ relation is real and an indicator of the value of coupling the increased metallicity sensitivity of the $hk$ index with the reddening and metallicity-independent H$\beta$ index and the minimal temperature-dependence of the [Fe/H] calibration based upon H$\beta$. The unweighted average of the four determinations is 
[Fe/H] $= -0.39 \pm 0.06$, while inclusion of a weight based upon the inverse of the standard error of the mean raises the average to [Fe/H] $= -0.37 \pm 0.05$. Though the difference between the two is small, the weighted average should be preferred. Over and above the larger internal scatter for the $m_1$ determinations, if comparable systematic errors are present within the zero-points of the calibrated photometry, the impact on the [Fe/H] estimate will be two to three times larger for $m_1$ than $hk$. Thus, the modest difference between the [Fe/H] derived from $m_1$ and that from $hk$ could be totally explained by a zero-point shift in the $m_1$ system of only 0.01 mag. In contrast, the reddening estimate would need to double to bring the two determinations into agreement. For $hk$, the $\Delta$[Fe/H]/$\Delta$$E(B-V)$ is less than 0.5.

\subsection{Comparison to Previous Determinations: Reddening}
For the reddening determination, only one additional approach has become available since the discussion in TAAT where $E(B-V)$ = 0.05 was adopted. The reddening maps of \citet{sc98} indicate $E(B-V)$ = 0.04 in the direction of NGC 2420, an upper limit along this line of sight but undoubtedly close to the cluster value given the significant distance of the cluster above the galactic plane. In short, the value we derive directly from the stars at the turnoff is consistent within the uncertainties with the commonly adopted range of $E(B-V)$ = 0.04 to 0.05 and adds support to the reliability of the DDO photometric technique applied in TAAT whenever possible.  

\section{DISCUSSION}
\subsection{The Metallicity}
The primary motivation for this investigation is the desire to place NGC 2420 on the reliably defined photometric system of abundances tied to the extended Str\"{o}mgren system, while testing recent suggestions that the cluster may have an overall abundance closer to solar. The photometric data for the stars at the turnoff clearly indicate an abundance half that of the sun using either the weighted average of the abundances based upon $m_1$ and $hk$ or the individual photometric values.  While the [Fe/H] derived from $hk$ is somewhat higher than the result implied from
$m_1$, the modest photometric errors and the high sensitivity of $hk$ to metallicity changes preclude a reconciliation with a higher abundance.  

Since the work of \citet{tab} and TAAT, no new independent abundance estimates have been generated for NGC 2420 with the exception of \citet{del}. In TAAT, the weighted average of the DDO and low resolution spectroscopy data was [Fe/H] = --0.27 $\pm$0.02 (s.e.m.) from 19 stars; DDO alone produced [Fe/H] = --0.28 $\pm$0.02 (s.e.m.) from 10 stars. However, the earlier work of \citet{fj}, a significant component of the data base in TAAT, has been expanded to more stars within previously studied clusters and to more clusters, and recalibrated by \citet{fj02}. The more recent work includes low resolution spectroscopy for 20 stars and finds [Fe/H] = --0.38 $\pm$0.02. While this result appears identical to that of the main sequence photometry, it must be remembered that the metallicity scales are not the same. In particular, the metallicity scale of \citet{fj02} differs from that of both \citet{fj} and TAAT. To reliably compare the results, we need to place the newer data on the original system of TAAT. 

Before beginning the transfer, two points need to be made. Even a quick glance at the final data of \citet{fj02} indicates that a problem exists with at least the zero-point of the scale. Of 39 clusters, only NGC 6791 at [Fe/H] = +0.11 has a metallicity of solar or above; 37 clusters have [Fe/H] = --0.10 or lower. On the system of TAAT, 62 clusters interior to $R_GC$ = 10 kpc have an average [Fe/H] of solar with a dispersion below 0.1 dex. While the average age of the clusters in \citet{fj02} is higher, this has no bearing on the question given the well established lack of an age-metallicity relation among open clusters  and field stars younger than 5 Gyrs starting with \citet{hi78} and \citet{tw80} and repeated many times since then, including \citet{fr95, fj02} and \citet{ed93}. Second, to emphasize a point already made in \citet{fj02}, while the adopted metallicity scale may require a zero-point and/or scale change, within the errors, the relative rankings of the clusters should be correct. Thus, patterns based upon the trends with metallicity, e.g., the existence of a galactic abundance gradient, should remain intact, though the slope may change.

How do the abundances from \citet{fj02} compare with those in TAAT? This apparently simple question has a somewhat complicated answer due to the method by which the spectroscopic scale was established. The overall calibration made use of a mixture of field giants and clusters. The metallicity scale of the field stars was based upon abundances defined using a mixture of high dispersion spectroscopy and DDO photometry. For the four clusters, the abundances were fixed at [Fe/H] = --0.1, --0.24, --0.42, and --0.79 for M67, NGC 7789, NGC 2420 and M71, respectively. The final calibration produces abundances indicating that the clusters and field stars are on the same system; the final open cluster values are --0.15, --0.24, and --0.38 for M67, NGC 7789, and NGC 2420, respectively. The revised scale has the effect of compressing the metallicity range of the entire sample compared to \citet{fj} since the abundances at the metal-rich end are reduced by approximately 0.1 dex while a slight increase occurs at the metal-poor end. The need for a compression of the scale was noted in TAAT. However, the appropriate change was found to be non-linear and consistent with the pattern established for converting from the older DDO scale to the revised one. For stars with [Fe/H]$_{FJ}$ below --0.20, a straight offset of 0.14 dex was added. For more metal-rich clusters, a linear transformation of [Fe/H] = 0.55[Fe/H]$_{FJ}$ + 0.05 was found, heavily influenced by the comparisons with \citet{pca}.

As a simple first attempt, we will calculate the offset between the DDO abundances of TAAT and the overlapping clusters in \citet{fj02}. In the original comparison, the common sample contained 14 clusters; the expanded sample contains 16. The average difference in [Fe/H], in the sense [DDO - SPE], is +0.09, with a standard deviation of only $\pm$0.08. The predicted dispersion based upon adding the quoted standard error of the mean for each cluster in quadrature is $\pm$0.06. 

A more complicated approach in an attempt to account for a potential non-linear offset can be achieved by transforming the revised data of \citet{fj02} to the older scale and applying the same transformation used in TAAT. There are 28 clusters common to the two spectroscopic surveys. Of these, three (Be 21, To 2, NGC 2112) are excluded from the comparison because of issues involving reddening, membership, and small number statistics in the original survey. For the remaining 25 clusters, a linear fit through the data gives [Fe/H]$_{old}$ = 1.33 [Fe/H]$_{new}$ + 0.10, with a scatter about the relation of $\pm$0.05 dex. Comparing the revised data to the transfer relation originally used to convert the spectroscopic data to the DDO system results in a modification of the zero-point by --0.04 to guarantee an average residual in [Fe/H] of 0.00 for the 16 clusters common to the two surveys. More importantly, the scatter among the residuals is reduced to $\pm$0.07 dex, a minor improvement compared to a straight offset but, again, only slightly larger than predicted from the uncertainty in the individual abundances alone. This confirms the point originally made in TAAT: the standard errors of the mean quoted for the individual cluster abundances are a reliable indicator of the true uncertainty in the estimate. For the sample as a whole, these consistently fall below 0.1 dex and, for over 40 clusters, the standard error of the mean is 0.05 dex or smaller.

Given the above, what does this mean for the abundance of NGC 2420? As already mentioned, the DDO estimate alone is [Fe/H] = --0.28. The revised [Fe/H] from \citet{fj02}, using a straight offset is --0.29; using the more complicated non-linear transfer gives [Fe/H] = --0.31. Taking all the techniques, DDO, $hk$, $m_1$, and spectroscopy into account leads to a weighted average of [Fe/H] = --0.32 $\pm$0.03 for the cluster, well below solar. 
  
\subsection{Low Resolution Reliability}
At this juncture in the discussion, the traditional next step would be an evaluation of the distance and age of the cluster through comparison with appropriate isochrones. Since the analysis of the metallicity and reddening confirms the values most often adopted in past studies and the exceptional definition of the CMD virtually guarantees that any new set of isochrones will be tested against the cluster, we will forego adding another redundant comparison between theory and observation. On the isochrone scale we normally adopt, the age of NGC 2420, as detailed in \citet{tab}, is 1.9 Gyr and, within the uncertainties, the combined slight increase in $E(B-V)$ and slight decrease in [Fe/H] leave the apparent distance modulus of 12.15 unchanged.

The above discussion of the link between the revised data of \citet{fj02} and TAAT does allow us to address the crucial question of the reliability of abundance patterns based upon photometry and/or low dispersion spectroscopy. This issue is important because of the periodic claims by some authors that the patterns discerned in TAAT are questionable because the sample is not homogeneous (see, e.g., \citet{ca98}), i.e., it is built using a mixture of photometry and low resolution spectroscopy. More recently, the implications have shifted such that the detailed patterns found in both TAAT and \citet{fj02} carry little weight because photometry and/or low resolution spectroscopy lead to relatively inaccurate results, thereby supplying partial justification for the use of high resolution spectroscopy, often of only 2 to 5 stars per cluster (see, e.g., \citet{car05}). Support for such claims is often tied to simplistic attempts to redefine the open cluster metallicity scale and/or the abundance gradient using merged samples with unnecessarily large final errors, as exemplified by the work of  \citet{gr00} (hereafter referred to as GR and discussed below) and \citet{ch03}, or using cluster parameters based upon CMD morphology incapable of providing the information desired at the level of accuracy required \citep{ha04}.

To understand the problem, we take a closer look at the attempt by GR to define an open cluster metallicity scale and, as a followup, to derive the slope of the galactic abundance gradient. Even simple techniques can have value as long as the conclusions they produce are consistent with the error bars generated by both the data and the analysis and those that use them are aware of the relative contribution of both. The basis of the discussion by GR is a composite sample of cluster abundances built upon the merging of data from high dispersion spectroscopy, low dispersion spectroscopy \citep{fj}, Washington, DDO, and UBV photometry. Since DDO and low resolution data were the foundation of TAAT, our review of the GR approach will be heavily weighted toward problems with these data, though we have no reason to believe that the problems we identify are limited to only these samples. Strangely, GR appears unaware of the revised and updated DDO data supplied in TAAT. In the sections that follow, the term $Friel$ refers to data from the entire low-resolution survey program defined by \citet{fj} and  
\citet{tho} and summarized in \citet{fr95}.

The first weakness in the derivation of a composite cluster sample and abundance scale 
is the lack of any attempt to update and homogenize the data, i.e., the data are simply lifted from a published table and adopted without modification, an issue common to a number of surveys, not just GR. While simple, it ensures the repetition of often avoidable errors, such as the inclusion of the incorrect metallicities for NGC 2112, discussed below, and Be 21, tied to an unreliable reddening estimate. Each cluster [Fe/H] on any given system is built from individual data points compiled assuming that the stars analyzed are single-star members of the cluster without any form of abundance anomalies, all of which have the same reddening as determined by some photometric and/or spectroscopic technique. However, the reddening adopted for a cluster can be different from one investigation to another, leading to an artificial scatter in the final average. Moreover, for individual clusters where only a handful of stars have been observed, the inclusion (or absence) of one or two stars can be critical. An extreme but important example of this is supplied by NGC 2112 where the average abundance from \citet{fj} is based upon a majority of stars which are non-members; the one probable member has a metallicity twice that of the original cluster average, as noted in TAAT and updated in \citet{fj02}. 
The value presented for $Friel$ in Table 2 of GR is approximately the incorrect original average, though this value and that from the Washington system are excluded in calculating the cluster mean, as is done for all clusters with high dispersion abundances. One may argue that such changes, while real, have little impact on the final result given the large intrinsic uncertainty in the cluster abundances, 0.15 dex for low dispersion spectroscopy and 0.15 to 0.20 dex for photometry. Unfortunately, these uncertainties quoted by GR are two to four times larger than the true errors for the DDO and low dispersion spectroscopy, as previously demonstrated.

The next issue is the definition of the cluster abundance scale, which is automatically assumed by GR to be that of the 26 high dispersion abundances garnered from the literature. While this could be a valid approach, the total absence of any information about the sources of the data, the means of analysis, the temperature scale, the reddening values, the number of stars included, or the means of combining abundances from multiple sources makes this an almost meaningless exercise. (We note that this analysis was part of a conference proceeding and publication space was undoubtedly limited.) This problem is compounded by the next step in the process, the transfer of all the data to the high dispersion spectroscopic system. Because of the small overlap (6 to 11 clusters) with the high dispersion data, GR calculates and applies a simple offset for each system. While this is the statistically justifiable approach given the small numbers, it clearly misses the point that transformations between one abundance system and another are unlikely to be that simple and use of such offsets can create unnecessary scatter. 

Given that these are valid points, one might question if they have any significant impact of the conclusions of GR? As the scatter in the offsets calculated by GR demonstrates, the answer is a definite yes. Before detailing the points, some issues for which we have no explanation must be noted. In the derivation of the offsets, GR finds an overlap of 6, 11, 9, and 11 clusters for $Friel$, Washington, UBV, and DDO, respectively. However, a check of the table shows the overlap to be 9, 11, 11, and 12. For the low resolution spectroscopy, the comparison is even more confusing. GR omits the results of \citet{tho} for six additional clusters, all of which were included in \citet{fr95}, two of which overlap with his spectroscopic data. Moreover, the abundances listed by GR for $Friel$ are not those found in \citet{fj} or \citet{fr95} and no explanation is supplied for the changes.

Returning to the issue of the scatter among residuals in calculating offsets, for the low dispersion spectroscopic technique, the rms scatter is given as 0.19 dex, while the DDO dispersion is 0.12 dex. Though this is consistent at a modest level with the claimed accuracy of these techniques by GR, the TAAT analysis and the comparisons in the previous section demonstrate that they are simply not plausible unless a significant error is present in the sample of GR. First, what is the real dispersion in the residuals for the GR data compared to $Friel$? If one uses all 9 clusters with overlapping data in Table 2 of GR, the mean offset, in the sense (FJ-HRS) is --0.14 $\pm$0.19 dex, the same value listed for 6 clusters by GR. If one uses the data as published by \citet{fj}, the same offset becomes --0.16 $\pm$0.13 dex. Not only is the dispersion significantly smaller, but a large portion of it comes from one cluster, NGC 2112, discussed earlier.  If this single point is dropped, the offset for the remaining 8 clusters is --0.13 $\pm$0.10. 

The fact that the abundance errors assumed by GR to be typical of photometric approaches (0.15 to 0.2 dex) and low resolution spectroscopy (0.15 dex) are exaggerated is easily demonstrated by one basic fact in TAAT. Without any correction for potential gradients with position or variation in abundance with age, 62 clusters interior to $R_{GC}$ = 10 kpc exhibit a dispersion in [Fe/H] based upon a mixed sample of DDO photometry and/or low dispersion spectroscopy of only $\pm$0.09 dex. In short, while the simplistic approach of GR creates a sample of 109 clusters compared to the 76 of TAAT, the larger errors in the final sample ensure that the conclusions add little to the discussion of the open cluster system except a mixture of redundancy and confusion. Even if one includes the wrong abundance for Be 21, as did GR, the data of TAAT generate an abundance gradient with an uncertainty in the slope (0.008) almost half that of GR (0.015). The claim that the observational errors in [Fe/H] for most clusters still approach $\pm$0.2 dex is more a reflection of the flawed analysis than the real data.  Finally, as demonstrated by the simulations of galactic disk evolution found in \citet{sc01}, repeated derivations of the slope of the abundance gradient tied to a sample dominated by a galactocentric distance from 7 kpc to 13 kpc will never produce a value significantly different from --0.07 $\pm$0.02 since the metallicity scales adopted for various photometric and spectroscopic systems are commonly linked by a few key objects. Thus, while mergers of data from multiple samples can produce modest shifts in the zero-point of the gradient, the slope remains relatively unchanged and, within the errors, the same value found in the first quality study of the abundance gradient by \citet{ja79}.

\section{SUMMARY}
The fundamental parameters of the old, metal-deficient open cluster NGC 2420 have been derived by direct analysis of photometrically identified single stars near the turnoff of the cluster CMD. Continuing the pattern found in previous papers in this series, the reddening estimate of $E(B-V)$ = 0.050 $\pm$0.004, based upon a combination of intermediate and narrow-band photometry, is in excellent agreement with the traditionally adopted values from a variety of less direct approaches, while being considerably more reliable. The metal abundances from each of the indices, $m_1$ and $hk$, are in good agreement, producing [Fe/H] = --0.37, though the errors for the abundance based upon $hk$ as defined relative to H$\beta$ are considerably reduced compared to those from $m_1$, a reflection of the higher sensitivity to changes in [Fe/H] for $hk$ compared to $m_1$ with data of the same photometric accuracy. The $hk$-based [Fe/H] of --0.32 is in excellent agreement with the independently derived [Fe/H] estimates from DDO and from the transformed low resolution abundances of \citet{fj02}. The universal agreement is that NGC 2420 has approximately half the metal abundance of the sun; we can see no means by which all of the above techniques can be collectively altered to render an abundance closer to solar.

Since the [Fe/H] for NGC 2420 is virtually the same as that derived in TAAT, the current result has no effect on the overall discussion except to confirm once again the reliability of the abundances based upon the giants (Papers I, IV, and V). The simple transformation of the data of \citet{fj02} to the metallicity scale of TAAT, however, does reaffirm the unique nature of the galactic disk near R$_{GC}$ = 10 kpc. The sample of \citet{fj02} contains 19 clusters interior to this point; on the TAAT scale only two have [Fe/H] $\leq$ --0.20. The abundance for NGC 2112 (--0.26) is based on only one star with an uncertainty greater than 0.1 dex while NGC 3960 (--0.24) suffers from highly variable reddening. In contrast, of the 14 clusters between R$_{GC}$ = 10 and 12 kpc, 10 have [Fe/H] below --0.20 and 6 of these are below --0.30.

Finally, the agreement among the results supplied by intermediate-band photometry of main sequence stars ($uvbyCa$H$\beta$) and red giants (DDO) and low resolution spectroscopy of red giants supports the effectiveness of all three techniques when the data are carefully obtained and the resulting abundances are consistently and reliably transformed to a common metallicity scale. In contrast, while it is the unique purview of high dispersion spectroscopy to supply individual elemental abundances, the promise of consistently reliable [M/H] estimates on a scale common to all observers remains elusive, as exemplified by the abundances for a cluster observed specifically by different groups to shed light on the galactic abundance gradient. For Cr 261, \citet{car05} find [Fe/H] $= -0.03 \pm 0.03$ based upon 5 red giants; from 4 red giants, \citet{fr03} find [Fe/H] $= -0.22 \pm 0.05$. The low resolution abundance from \citet{fj02}, transformed to the DDO scale of TAAT, is [Fe/H] = --0.05 $\pm$0.03.

\acknowledgements
Misty M. Cracraft and Delora C. Tanner participated in the observations and analysis, each contributing component software used in the reduction of these data.  Their software and this analysis were completed in partial fulfillment for their master's degrees at the University of Kansas. The data used in this project would not have been accessible without the help of Con Deliyannis and the excellent support provided by the WIYN 0.9m staff at KPNO. We also gratefully acknowledge Con's thoughtful comments on the manuscript. Extensive use was made of the SIMBAD database, operating at CDS, Strasbourg, France  and the WEBDA database maintained at the University of Geneva, Switzerland. 
The cluster project has been helped by support supplied through the General Research Fund of the University of Kansas and from the Department of Physics and Astronomy.

\clearpage

\figcaption[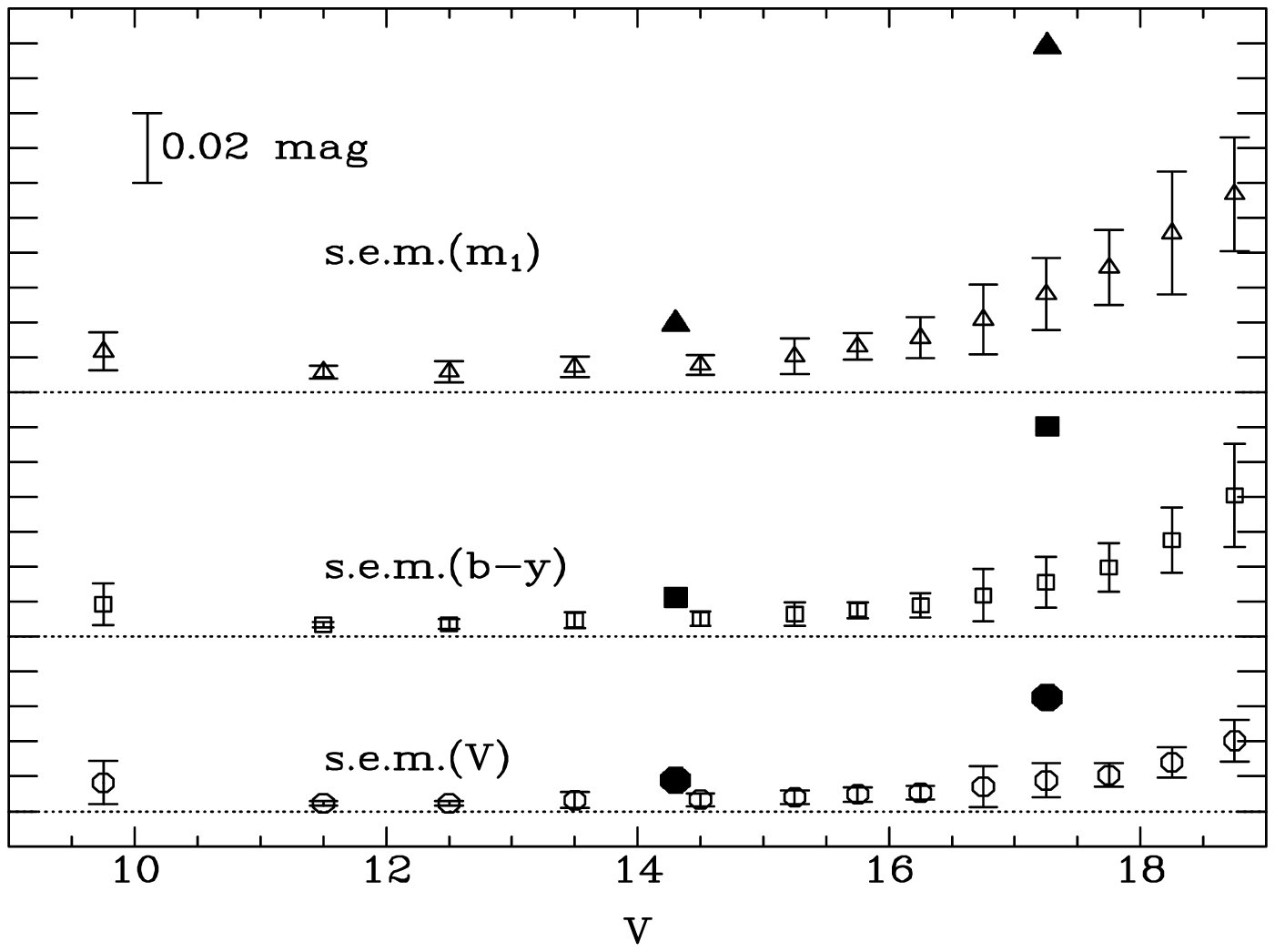]{Standard errors of the mean (s.e.m.) for the $V$, $b-y$, and $m_1$ as a function of $V$. Open symbols show the average s.e.m. while the error bars denote one sigma dispersion. Filled symbols are stars identified as potential variables. \label{fig1}} 

\figcaption[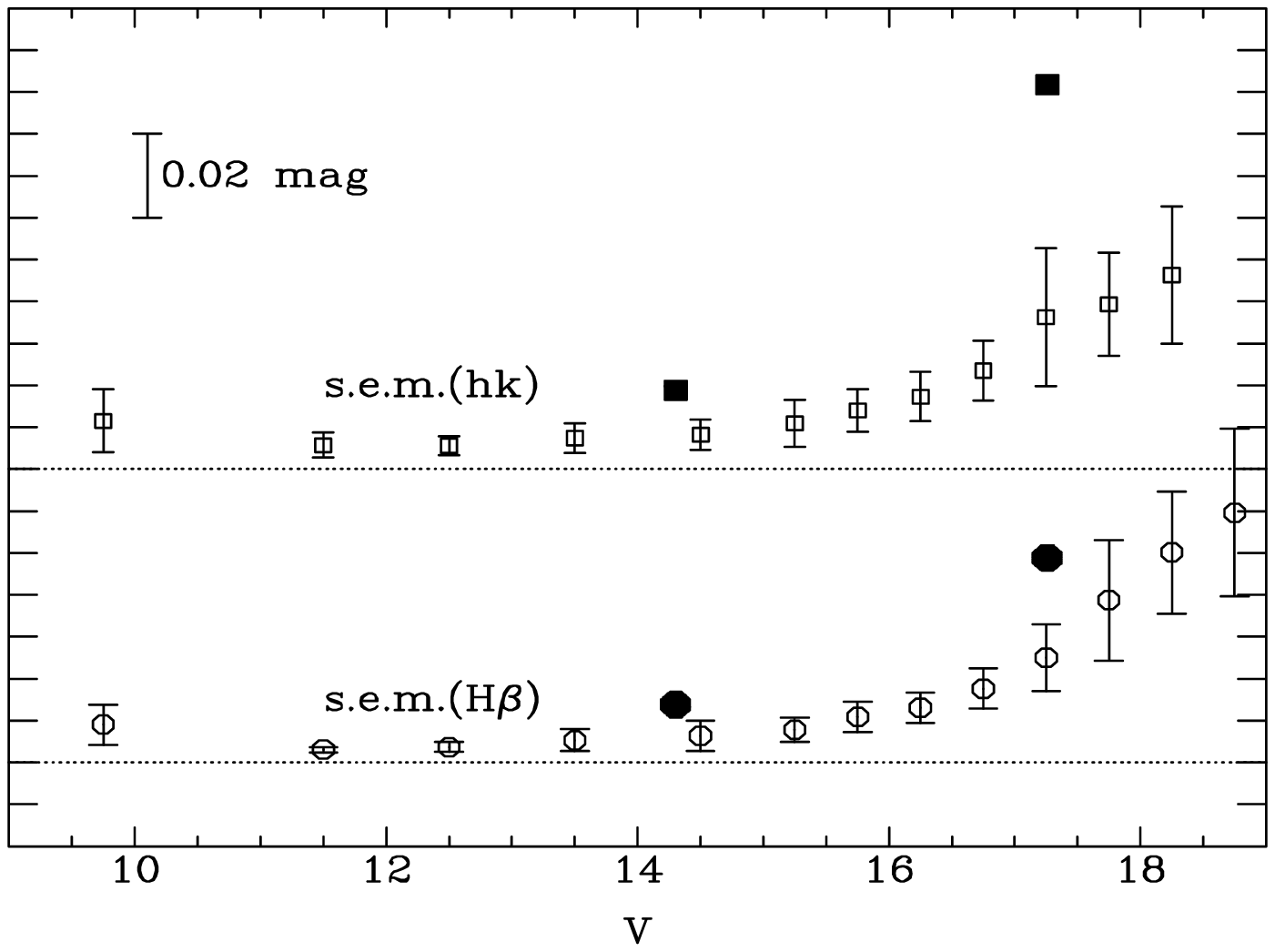]{Same as Fig. 1 for the $hk$, and H$\beta$ indices. \label{fig2}}

\figcaption[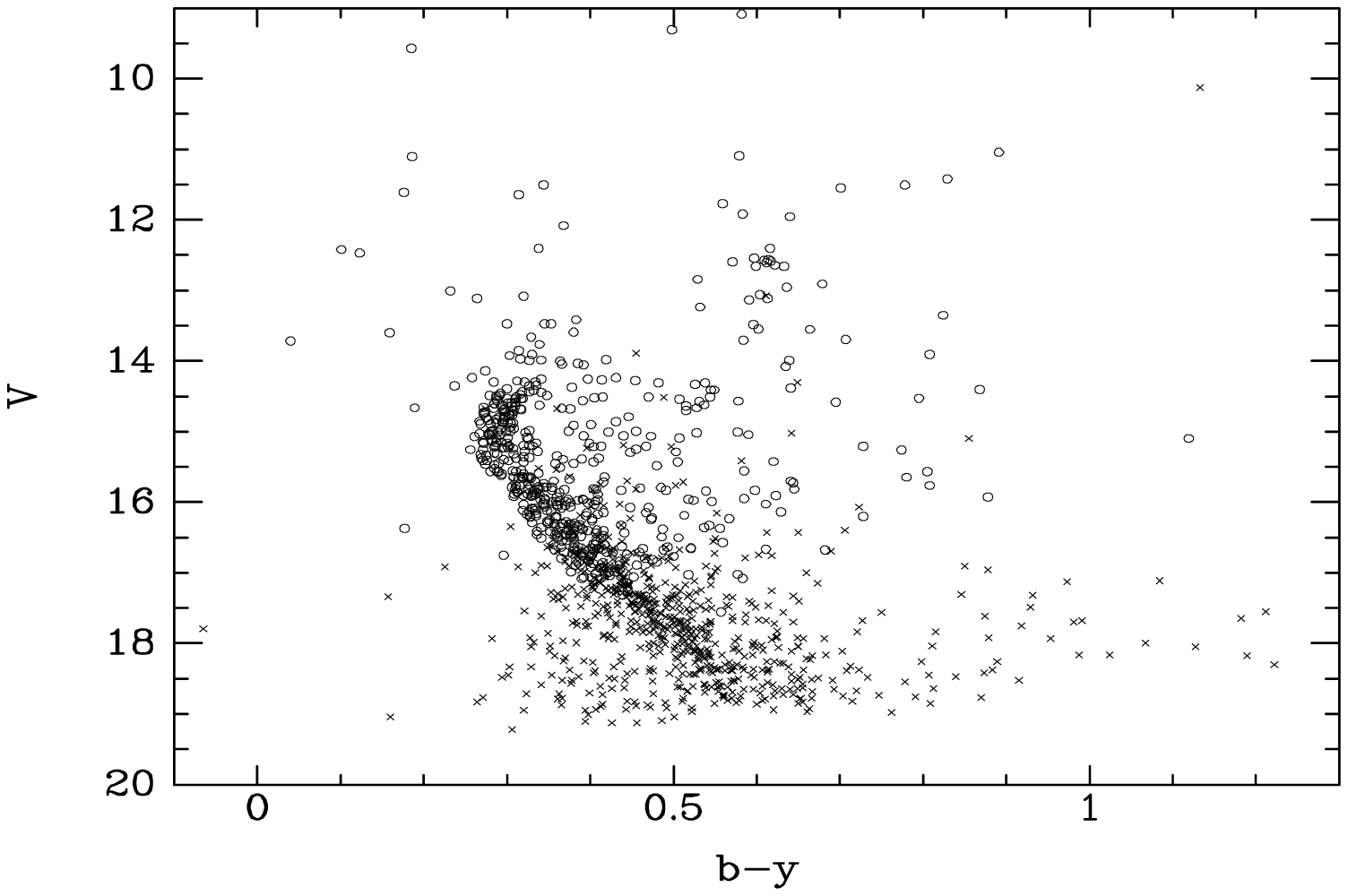]{Color-magnitude diagram for stars with at least 2 observations each in $b$ and $y$. Crosses are stars with internal errors in $b-y$ greater than 0.010 mag.\label{fig3}}

\figcaption[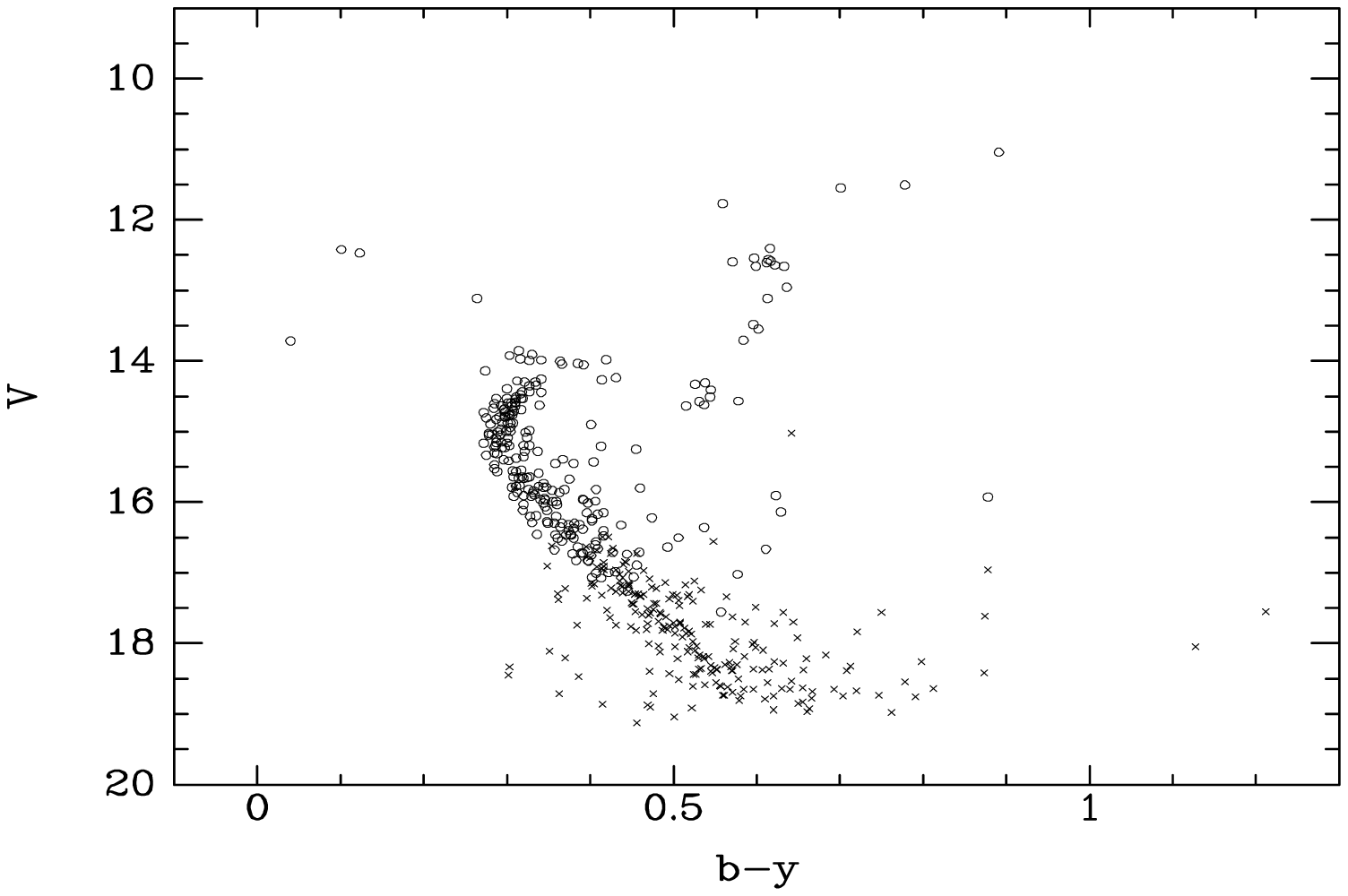] {Same as Fig. 3 for stars within 300 pixels of the cluster center. \label{fig4}}

\figcaption[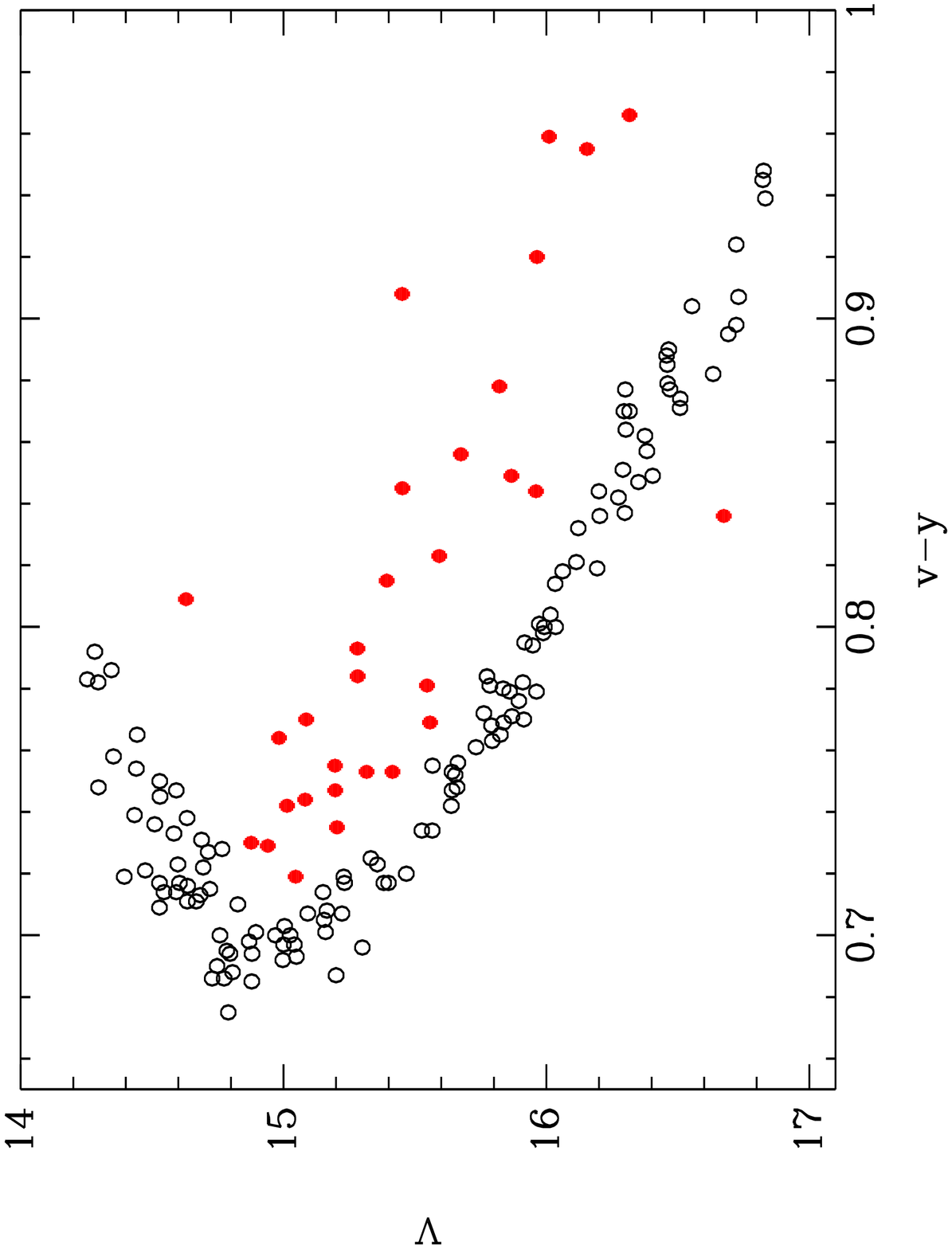] {$V,v-y$ CMD for cluster core stars at the turnoff. Filled red circles are stars tagged as potential binaries, nonmembers, or photometric anomalies. \label{fig5}}

\figcaption[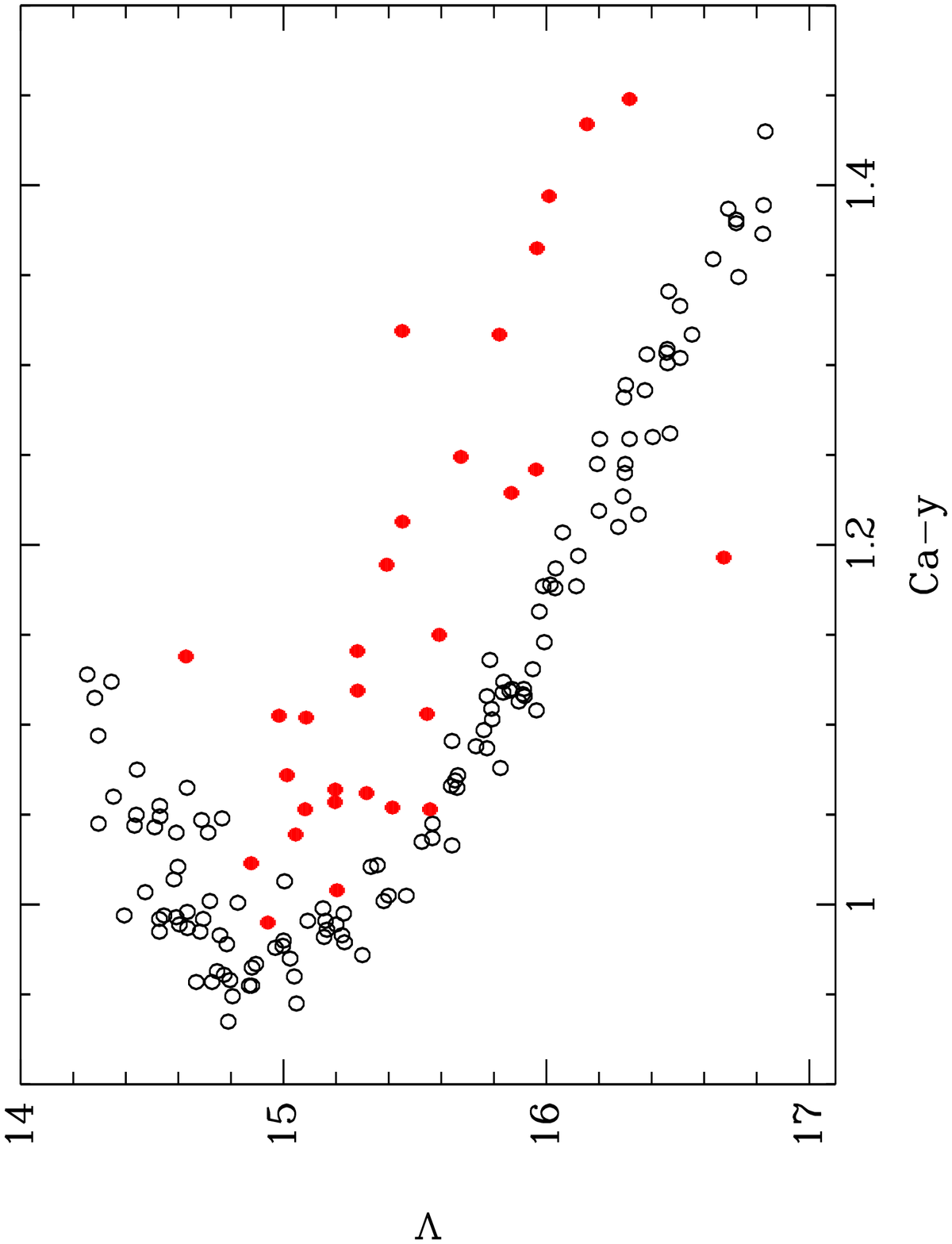] {$V,Ca-y$ CMD for cluster core stars at the turnoff. Symbols have the same meaning as in Fig. 5. \label{fig6}}

\figcaption[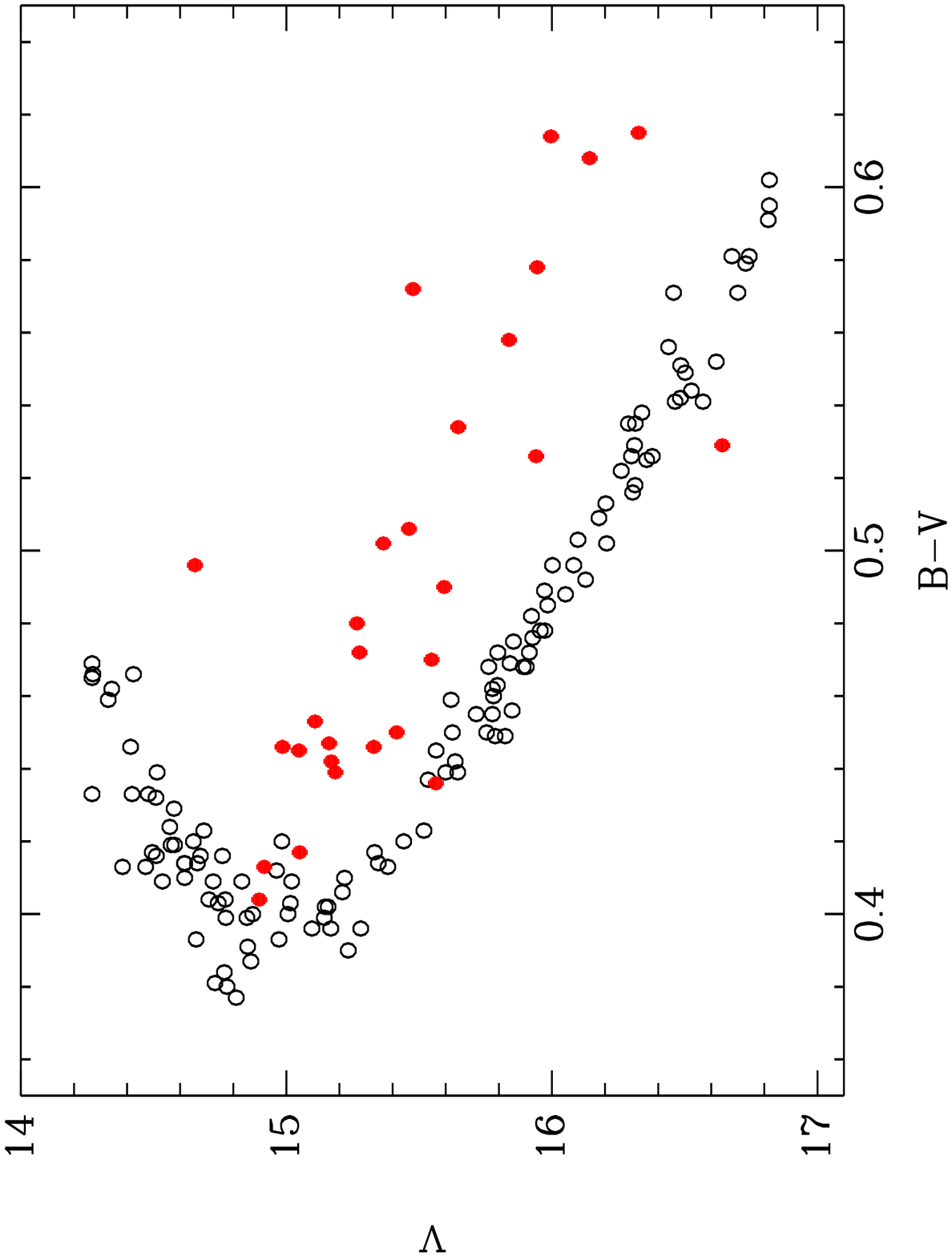]{Traditional CMD for the same stars as Fig. 5.\label{fig7}}

\newpage
\plotone{f1.eps}
\newpage
\plotone{f2.eps}
\newpage
\plotone{f3.eps}
\newpage
\plotone{f4.eps}
\newpage
\includegraphics[scale=0.53,angle=-90]{f5.eps}
\newpage
\includegraphics[scale=0.53,angle=-90]{f6.eps}
\newpage
\includegraphics[scale=0.53,angle=-90]{f7.eps}
\newpage
\begin{deluxetable}{rrrrrrrrrrrrrc}
\tabletypesize\scriptsize
\tablenum{1}
\tablecolumns{14}
\tablewidth{0pc}
\tablecaption{Extended Str\"omgren Photometry in NGC 2420}
\tablehead{
\colhead{ID}     & 
\colhead{X}     & 
\colhead{Y}     & 
\colhead{$V$}     & 
\colhead{$b-y$}     & 
\colhead{$m_1$}     & 
\colhead{H$\beta$}     & 
\colhead{$hk$}     & 
\colhead{$\sigma_V$}     & 
\colhead{$\sigma_{by}$}     & 
\colhead{$\sigma_{m1}$}     & 
\colhead{$\sigma_{\beta}$}     & 
\colhead{$\sigma_{hk}$}     & 
\colhead{$N_{ybvwnCa}$}    }
\startdata
    82& -66.72&-458.62&  9.087& 0.582& 0.427&\nodata& 1.156& 0.004&0.006&0.009&\nodata&0.007&  4,2,9,0,5,11 \cr
  6001&  20.43& 594.64&  9.309& 0.498& 0.384&\nodata& 1.024& 0.008&0.009&0.012&\nodata&0.010&  4,2,9,0,8,10 \cr
   244& 218.58& 362.93&  9.568& 0.184& 0.227 &2.742& 0.488& 0.004&0.005&0.007&0.006&0.006&  4,2,3,1,6,9 \cr
  6002&-594.76& 492.81& 10.123& 1.133& 0.454 &2.794& 1.780& 0.017&0.018&0.019&0.012&0.023&  2,2,8,4,4,2 \cr
  6003&-127.26&  12.07& 11.045& 0.891& 0.674 &2.545& 1.810& 0.002&0.004&0.005&0.002&0.005& 16,8,11,16,11,13 \cr
\enddata
\tablecomments{An abbreviated form of Table 1 is presented here to illustrate format; the full data table will appear in the Astronomical Journal or may be requested from the authors.}
\end{deluxetable}


\begin{deluxetable}{lcccl}
\tablenum{2}
\tablecolumns{5}
\tablewidth{0pc}
\tablecaption{Summary of Comparisons with Other Photometric Surveys}
\tablehead{
\colhead{Ref.} & \colhead{$\Delta V$} & \colhead{Std.Dev.}   &
\colhead{N} & \colhead{Class} }
\startdata
 Anthony-Twarog et al. (1990)   &    0.001  &    0.034 & 531 & ALL \cr
     &    0.001  &    0.030 & 521 & $\Delta V \leq 0.10$\cr
       &    0.005  &    0.022 & 229 & $V \leq 16.5$ \cr
            &    0.005  &    0.021 & 228 & $\Delta V \leq 0.075$\cr
  Stetson (2000)                &    0.007  &    0.027 & 177 & ALL \cr
        &    0.000  &    0.023 & 177 & $\Delta V = -0.111(B-V)+ 0.046$\cr
              &    0.017  &    0.023 & 114 & $V \leq 16.5$\cr
                    &    0.017  &    0.022 & 113 & $\Delta V \leq 0.075$ \cr   
\enddata
\tablecomments{$\Delta V$ is defined in the sense (Table 1 - references).}
\end{deluxetable}

\enddocument